\documentclass[final,3p,times,authoryear]{elsarticle}
\usepackage{epsfig}
\usepackage{amssymb}
\usepackage{amsthm}
\usepackage{hyperref}
\usepackage{ragged2e}
\usepackage[table]{xcolor}
\usepackage{booktabs}
\usepackage{etoolbox}
\usepackage{lipsum}
\usepackage{sidecap}
\graphicspath{ {./fig/} }
\usepackage{amsfonts} 
\usepackage{amsmath}
\usepackage{amsthm}
\usepackage{ulem}
\usepackage[table]{xcolor}
\usepackage{graphicx} % Allows including images
\journal{Spatial Statistics}

\begin{document}

\begin{frontmatter}

%% Title, authors and addresses

%% use the tnoteref command within \title for footnotes;
%% use the tnotetext command for theassociated footnote;
%% use the fnref command within \author or \affiliation for footnotes;
%% use the fntext command for theassociated footnote;
%% use the corref command within \author for corresponding author footnotes;
%% use the cortext command for theassociated footnote;
%% use the ead command for the email address,
%% and the form \ead[url] for the home page:
%% \title{Title\tnoteref{label1}}
%% \tnotetext[label1]{}
%% \author{Name\corref{cor1}\fnref{label2}}
%% \ead{email address}
%% \ead[url]{home page}
%% \fntext[label2]{}
%% \cortext[cor1]{}
%% \affiliation{organization={},
%%            addressline={}, 
%%            city={},
%%            postcode={}, 
%%            state={},
%%            country={}}
%% \fntext[label3]{}
\title{Accounting for Multiple Covariates in Non-Stationary Geostatistical Modelling}
%% use optional labels to link authors explicitly to addresses:

%% \author[label1,label2]{}
%% \affiliation[label1]{organization={},
%%             addressline={},
%%             city={},
%%             postcode={},
%%             state={},
%%             country={}}
%%
%% \affiliation[label2]{organization={},
%%             addressline={},
%%             city={},
%%             postcode={},
%%             state={},
%%             country={}}

\author[inst1]{Olatunji Johnson}
\affiliation[inst1]{organization={Department of Mathematics},
            addressline={University of Manchester}, 
            city={Manchester},
            country={UK}}
\ead{olatunji.johnson@manchester.ac.uk}
\author[inst2]{Bedilu A Ejigu}
\affiliation[inst2]{organization={Department of Statistics, },%Department and Organization
            addressline={Addis Ababa University}, 
            city={Addis Ababa},
            country={Ethiopia}}
            \ead{bedilu.alamirie@aau.edu.et}
\author[inst3]{Ezra Gayawan}
\affiliation[inst3]{organization={Department of Statistics, },%Department and Organization
            addressline={Federal University of Technology,}, 
            city={Akure},
            country={Nigeria}}
            \ead{egayawan@futa.edu.ng}
\begin{abstract}
Model-based geostatistics (MBG) is a subfield of spatial statistics focused on predicting spatially continuous phenomena using data collected at discrete locations. Geostatistical models often rely on the assumptions of stationarity and isotropy for practical and conceptual simplicity. However, an alternative perspective involves considering non-stationarity, where statistical characteristics vary across the study area. While previous work has explored non-stationary processes, particularly those leveraging covariate information to address non-stationarity, this research expands upon these concepts by incorporating multiple covariates and proposing different ways for constructing non-stationary processes. Through a simulation study, the significance of selecting the appropriate non-stationary process is demonstrated. The proposed approach is then applied to analyse malaria prevalence data in Mozambique, showcasing its practical utility. 
\end{abstract}

\end{frontmatter}
%% \linenumbers

% \textcolor{Blue}{Possible list of journals that we may consider:
% \begin{enumerate}
% \item Spatial statistics
% \item Journal of the Royal Statistical Society, Series C
% \item Statistics in Biosciences
% \item International journal of health geographics
% \end{enumerate}}

\section{Introduction}

Model-based geostatistics (MBG) \citep{diggle1998model}, is a field within spatial statistics that offers valuable tools for making spatially continuous inferences. It specifically allows us to predict spatially continuous phenomena based on data collected at discrete locations within a defined region of interest. MBG operates under a principled likelihood-based approach, leveraging the "first law of geography," which suggests that proximity implies a higher level of spatial correlation. This enables geostatistical models to effectively borrow information across space to estimate values at any location within the studied area. Given its adaptability to low-resource settings where disease registries may be lacking, MBG has found increasing applications in epidemiological studies conducted in developing countries. Notably, MBG has been instrumental in mapping various infectious diseases, including malaria \citep{ejigu2023new}, Loa loa \citep{johnson2022geostatistical}, trachoma \citep{amoah2022model}, soil-transmitted helminths \citep{mogaji2022estimating}, and onchocerciasis. These applications have provided valuable insights for monitoring disease burden using data from household surveys.

Let $Y$, denote the outcome measured at a specific location $x$. The standard linear geostatistical model for the outcome $Y$ at location $x$ takes the form:

\begin{equation}
\label{eq:standardGeo}
Y_i = d^\top(x_i) \beta +  S(x_i) + Z_i, \quad \quad \text{for} ~ i = 1, \ldots, n 
\end{equation}

where $d(x)$ is a vector of explanatory variables at location $x$ with associated regression coefficients $\beta$ and $Z_i$ is often referred to as \textit{nugget effect} and assumed to be a set of independent and identically distributed (i.i.d.) Gaussian random variables with mean zero and variance $\tau^2$. The common assumption on $S(x)$ is that it is stationary and isotropic implying that
\begin{equation*}
   Cov\{S(x), S(x')\}  = \sigma^2 \rho(u; \phi),
\end{equation*}
where $\sigma^2$ is the variance of the process $S$, $u = ||x-x'||$ is the Euclidean distance between locations $x$ and $x'$ and $\rho(u; \phi)$ is a correlation function with parameters $\phi$. We usually specify $\rho(u; \phi)$ to be a member of the Mat\'ern family \citep{matern1960spatial}.

% This assumption of stationarity and isotropy is common in geostatistical models, simplifying the modelling process by solely depending on the spatial distance between locations. The model allows us to explore the spatial correlation structure and make predictions for unobserved locations based on observed data.

The assumption of stationarity and isotropy is common in geostatistical models due to practical and conceptual reasons. Stationarity and isotropy use universal covariance functions that are solely a function of distance to simplify the model specification and modelling process and allow for efficient estimation of model parameters. However, it is essential to recognise that these assumptions may not always hold in real-world scenarios. As an alternative, we can assume non-stationarity, where statistical properties vary across the study area. In this case, the mean and/or variance of the process can vary spatially based on covariates or other factors. These are particularly useful when there are evident spatial trends or gradients in the data.

%%% review of non-stationary process
Several models for non-stationary processes have been proposed in the past \citep{Sampson1992, higdon1998process, Paciorek2006, Schmidt2011}. The space deformation methods proposed by \cite{Sampson1992} is a pioneering paper in nonstationary covariance function modelling.  In their proposal, the basic idea is to transform the geographic region $D$ to a new region $G$, a region such that stationarity and isotropy hold on $G$. However, the space deformation method suffers from two limitations: i) it cannot quantify the uncertainty introduced in estimating the mapping from $D$ to $G$, and ii) the estimated mapping is often not one-to-one and folded over itself.

The process convolution method \citep{higdon1998process} is the most popular method of constructing a nonstationary process because it is easier to specify using the kernel function rather than the covariance function. This approach is based on the convolution of a stationary process with a nonstationary kernel to obtain a nonstationary process. However, this approach is highly parameterized and therefore difficult to fit, since it is hard to estimate nonstationary behaviour using only a single realization of a spatial process.

On the other hand, other approaches allow incorporating covariates in specifying the covariance structure of a spatial process \citep{Schmidt2011, ejigu2020geostatistical}. These approaches consider covariate information to handle non-stationarity in different ways.  The method proposed by \citep{Schmidt2011} relax the assumption of stationary Gaussian process by accounting for covariate information in the covariance structure of the process to allow the latent space model of \cite{Sampson1992} to be of dimension $D>2$.  The method proposed by \citep{ejigu2020geostatistical} directly incorporates spatially referenced covariates into the covariance function but is confined to the inclusion of a single covariate. Considering that different parts of the spatial domain may be influenced by different sets of covariates, the model's flexibility to adapt locally is enhanced by incorporating multiple covariates. This allows the model to effectively capture diverse variations across the spatial domain.

%%% contribution of this paper
The main contribution of this paper is to propose an extension to the approach proposed by \cite{ejigu2020geostatistical}, by allowing for the inclusion of multiple covariates in the covariance of the spatial process. Specifically, we consider ways to combine multiple correlation functions, either by multiplying or adding them to give a valid corresponding correlation function.

%% structure of the paper
The subsequent sections of the paper are structured as follows. Section \ref{sec:method} provides a review of the existing approach and introduces our proposed extension. In Section \ref{sec:simulation}, we assess the robustness of our proposed approach via a simulation study. Section \ref{sec:application} showcases the application of the proposed approach using malaria prevalence data in Mozambique. Lastly, in Section \ref{sec:discussion}, we conclude with a discussion of the proposed method and explore potential avenues for further methodological extensions.

\section{Review of existing methods and the proposed method}
\label{sec:method}

\subsection*{Non-stationary geostatistical model with one covariate}
\cite{ejigu2020geostatistical} proposed a non-stationary geostatistical model with one covariate, in their approach,  they replace the stationary Gaussian process in Equation \eqref{eq:standardGeo} with another Gaussian process that is a function over both space $x$ and a covariate $e$, denoted as $S(x_i, e_i)$. The model then takes the form 
\begin{equation}
\label{eq:standardGeo2}
Y_i = d^\top(x_i) \beta +  S(x_i, e_i) + Z_i,
\end{equation}
They assumed this to be a Gaussian process with mean zero and covariance function 
\begin{equation*}
   Cov\{S(x, e), S(x', e')\}  = \sigma^2 \rho(x, x'; e, e'; \phi).
\end{equation*}
And then assumed a separable correlation function in which the space-covariate covariance function was decomposed into the product of a purely spatial and a purely covariate-dependent covariance function such that $\rho(x, x'; e, e'; \phi) = \rho_1(x, x'; \phi_1) \rho_2(e, e'; \phi_2)$, where $\rho_1(x, x'; \phi_1)$ is the spatial correlation function, and $\rho_2(e, e'; \phi_2)$ is the correlation function associated with covariate. They noted that the covariate $e$ can be included in the fixed part of the model as well as in the covariance function. However, in a realistic situation, most spatial data are accompanied by more than a covariate and the desire is to investigate the influence of these covariates on the spatial outcome. Consequently, it becomes necessary to extend the proposed method to include more covariates.

\subsection*{Non-stationary geostatistical model with more than one covariate}
Here we proposed an extension to the method above by allowing for the inclusion of more than one covariate, which allows for the interaction between the spatial component and the covariates. This is essential for capturing local spatial heterogeneity and addressing possible complexities of the data. Given that different parts of the spatial domain may be influenced by different sets of covariates, the utilization of multiple covariates enables the model to adapt locally, effectively capturing diverse influences in different regions. For convenience's sake, we restrict ourselves to two covariates and later show how this can be generalised to more than two covariates. Then we proposed the stochastic process $S(x, e, t)$, where $e$ and $t$ are covariates. $S(x, e, t)$ can be modelled as a zero mean Gaussian process with a covariance function
\begin{equation*}
    Cov\{S(x, e, t), S(x', e', t')\}  = \sigma^2 \rho(x, x'; e, e'; t, t'; \phi).
\end{equation*}

We consider three decompositions of the covariance function as follows:

\begin{equation}
\label{eq:cov1}
    Cov\{S(x, e, t), S(x', e', t')\}  = \sigma^2 \rho_1(x, x'; \phi_1)\rho_2(e, e'; \phi_2) \rho_3(t, t'; \phi_3)
\end{equation}

\begin{equation}
\label{eq:cov2}
    Cov\{S(x, e, t), S(x', e', t')\}  = \sigma^2 (\rho_1(x, x'; \phi_1) \rho_2(e, e'; \phi_2) + \rho_1(x, x'; \phi_1) \rho_3(t, t'; \phi_3)),
\end{equation}

and 
\begin{equation}
\label{eq:cov3}
    Cov\{S(x, e, t), S(x', e', t')\}  = \sigma^2 (\rho_1(x, x'; \phi_1) + \rho_2(e, e'; \phi_2) + \rho_3(t, t'; \phi_3)),
\end{equation}

where $\rho_1(x, x'; \phi_1)$ is the spatial correlation function,  $\rho_2(e, e'; \phi_2)$ is the correlation function associated with covariate $e$, and $\rho_3(t, t'; \phi_3)$ is the correlation function associated with covariate $t$. The aforementioned covariance structures presented in  Equations \eqref{eq:cov1} \eqref{eq:cov2} and \eqref{eq:cov3} are equivalent to the covariance structure of $S(\cdot)$ in models named Model 1, Model 2, and Model 3, respectively:

\textbf{Model 1}

\begin{equation}
\label{eq:model1}
    Y_i = d^\top(x_i) \beta +  S(x_i, e_i, t_i) + Z_i,
\end{equation}

\textbf{Model 2}

\begin{equation}
\label{eq:model2}
    Y_i = d^\top(x_i) \beta +   S(x_i, e_i) + S(x_i, t_i)  + Z_i,
\end{equation}

\textbf{Model 3}

\begin{equation}
\label{eq:model3}
    Y_i = d^\top(x_i) \beta +  S(x_i) + S(e_i) + S(t_i) + Z_i,
\end{equation}

% The process $S(x, e, t)$ can be decomposed as 
% \begin{equation}
%     S(x, e, t) = S(x) + S(e) + S(t),
% \end{equation}
% or 

% \begin{equation}
%     S(x, e, t) = S(x) S(e) S(t),
% \end{equation}
% where $S(x)$ is purely spatial stochastic process, and $S(e)$ and $S(t)$ are purely covariate stochastic process with associated covariates $e$ and $t$, respectively.

\par We assume that $\rho_1(x, x'; \phi_1)$ is a Matérn function \citep{matern1960spatial}, defined as 

\begin{equation*}
    \rho(u) =  \frac{1}{\Gamma(\kappa) 2^{\kappa -1}} (u/\phi)^\kappa K_\kappa (u/\phi)
\end{equation*}
where $K_\kappa (\cdot)$ denotes the modified Bessel function of the third kind of order $\kappa$, $u = \|x - x'\|$ is the Euclidean distance between location $x$ and $x'$, $u = |e - e'|$ is the absolute difference between the covariate $e$ and $e'$ or $u = |t - t'|$ is the absolute difference between the covariate $t$ and $t'$ and $\phi$ is the scale parameter which controls the rate at which the correlation gets close to zero with increasing separation distance $u$. According to \cite{zhang2004inconsistent}, it can be difficult to estimate $\kappa$ in practice as it requires a large amount of data. Therefore, we fixed the value of $\kappa$ to 1.5 in the simulation study which then corresponds to $\rho(u) = \left( 1 + \frac{\sqrt{3}u}{\phi}\right) \exp{\left( -\frac{\sqrt{3}u}{\phi}\right)}$. 

\subsection*{Inference: maximum likelihood estimation}
Maximum likelihood estimation (MLE) is a widely used statistical technique for parameter estimation in models. It aims to identify the parameter values within a probability distribution that yield the most optimal fit to the observed data. This is achieved by maximizing the likelihood function, which measures the extent to which the chosen distribution and its associated parameters can account for the observed data. In this case, the likelihood function is a multivariate normal distribution. Specifically, let $\theta = (\beta, \sigma^2, \phi_1, \phi_2, \phi_3, \tau^2)^\top$ denote the vector of the parameters and $y^\top = (y_1, \ldots, y_n)$ denotes the observed dataset, the log-likelihood function is given by 
\begin{equation*}
\ell(\theta) = \ln\left( \mathcal{L}({\theta} \right) = -\frac{n}{2} \ln(2\pi) - \frac{1}{2} \ln|{\Sigma}| - \frac{1}{2} (y - D \beta)^\top {\Sigma}^{-1} (y - D \beta),
\end{equation*}
where $D$ is an $n \times p$ matrix of explanatory variables and $\Sigma$ is the $n \times n$ covariance matrix. We used numerical optimization algorithms in R \citep{R} to find the maximum likelihood estimate $\hat{\theta}$. %\textcolor{red}{Here, I think it will be useful to be a bit more detailed on the nature of the algorithm and the specific package used in R.} \textcolor{blue}{The function used to fit our proposed model is not yet embedded in the CRAN available R packages. Mentioning the GitHub link for the function to  fit the model may be enough (currently it is mentioned under Discussion section, but it would be nice to mention it here). Once this paper published we may consider to document all necessary function and create a standalone package for this methodology}.

\section{Simulation Study}
\label{sec:simulation}

The purpose of the simulation study is twofold: To evaluate the effects of misspecifying the non-stationary covariance function on 1) the parameter estimation $\theta$, and 2) the spatial prediction of the outcome $Y$.

We consider three different data-generating mechanisms, referred to as Model 1, Model 2, and Model 3. In all datasets, we simulate $n=700$ observations and set the parameters to $\theta = (\beta_0, \beta_1, \beta_2, \sigma^2, \phi_1, \phi_2, \phi_3)^\top = (1, 0.5, -0.5, 0.5, 0.3, 0.2, 0.1)^\top$. The domain is defined as $[0,1] \times [0, 1]$, and sampling locations are simulated using a standard uniform distribution. Additionally, two predictors are simulated from the uniform distribution Unif[-1, 1]. Subsequently, $S$ is sampled from a multivariate normal distribution with a mean of zero and variances as specified in Equations \ref{eq:cov1}, \ref{eq:cov2}, and \ref{eq:cov3}, and Models 1, 2, and 3 are evaluated.

Three scenarios are created for the analysis:
\begin{itemize}
    \item In Scenario 1, data were simulated from Model 1, and the analysis was performed using Models 1, 2, and 3.
\item In Scenario 2, data were simulated from Model 2, and the analysis was performed using Models 1, 2, and 3.
\item In Scenario 3, data were simulated from Model 3, and the analysis was performed using Models 1, 2, and 3.
\end{itemize}
Therefore, Model 1 is correctly specified in Scenario 1, Model 2 is correctly specified in Scenario 2, and Model 3 is correctly specified in Scenario 3.

\subsection{Evaluating accuracy of the parameter estimate}

As suggested by \cite{burton2006design}, the accuracy of each parameter $\hat{\theta}_k$ can be assessed by calculating the percentage relative bias (PRB), given by $$ PRB =  \frac{1}{B} \sum_{k = 1}^{B} (\hat{\theta}_k - \theta_k)/\theta_k \times 100,$$ where $B$ is the number of simulations, and $\hat{\theta}_k$ and $\theta_k$ are the estimated and true values of the parameters, respectively. This analysis assumes that none of the parameters are equal to zero. PRB provides insight into the direction and magnitude of bias. A negative PRB indicates an underestimation, while a positive PRB indicates an overestimation.  

Additionally, the coverage of the 95\% Wald-type confidence interval was calculated, representing the proportion of times the interval contained the "true" performance value. The coverage probability is given as 
$$CP = \frac{1}{B}  \sum_{j = 1}^{B} I \left(\hat{\theta}_{k, lower}^{(j)} \leq \theta_k^{(j)}  \leq \hat{\theta}_{k, upper}^{(j)} \right), $$
where $\hat{\theta}_{k, lower}^{(j)}$ and $\hat{\theta}_{k, upper}^{(j)}$ are the lower and upper Wald type 95\% confidence interval.

\subsection{Evaluating the predictive performance of the model}
We assess the predictive model performance using three metrics: bias, root mean square error (RMSE), and coverage probability (CP). These metrics are defined as follows:
\begin{align}
    bias &= \frac{1}{nB} \sum_{i = 1}^{n} \sum_{j = 1}^{B} \left(\hat{Y}_i^{(j)} - Y_i^{(j)}\right), \notag \\
    RMSE &= \frac{1}{nB} \sum_{i = 1}^{n} \sum_{j = 1}^{B} \sqrt{\left(\hat{Y}_i^{(j)} - Y_i^{(j)}\right)}, \notag \\
    CP &= \frac{1}{nB} \sum_{i = 1}^{n} \sum_{j = 1}^{B} I \left( Y_i^{(j)} \in  PI_{0.95}^{(j)} \right), \notag
\end{align}

where $Y_i^{(j)}$ and $\hat{Y}_i^{(j)}$ are the true and predicted values of the outcome, respectively; $I \left( \hat{Y}_i^{(j)} \in  PI_{0.95}^{(j)} \right)$ is an indicator function that takes the value 1 if $Y_i^{(j)}$ is inside the 95\% prediction interval denoted by $PI_{0.95}^{(j)}$ and 0 otherwise. 

\subsection{Simulation result}
\begin{figure}[htp]
    \centering
    \includegraphics[scale = 0.5]{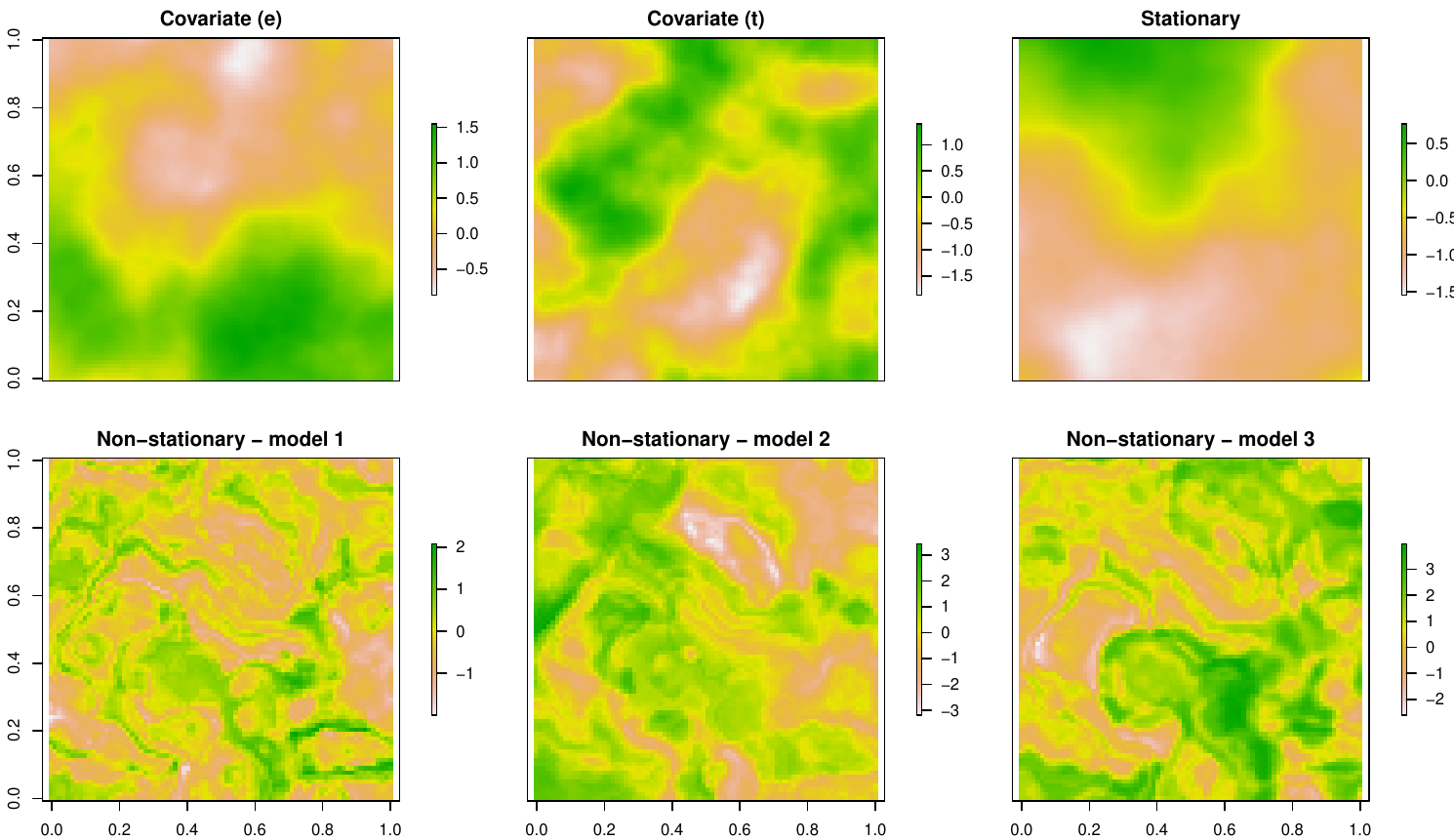}
    \caption{An example of a simulated surface of the spatial covariates $e$ (top left panel) and $t$ (top middle panel), stationary process (top right panel) and non-stationary process from models 1, 2 and 3 (lower panels). }
    \label{fig:process}
\end{figure}

To demonstrate the proposed process, Figures \ref{fig:process}, shows the simulated surface of the covariates $e$ and $t$, the stationary process $S(x)$ (if covariates were not included) and the resulting non-stationary process, for models 1, 2 and 3, respectively. The Figure was generated using the following parameters $(\sigma^2, \phi_1, \phi_2, \phi_3)^\top = (0.5, 0.3, 0.2, 0.1)^\top$. Clearly, for the non-stationary processes, the trend is not constant in space. 

Table \ref{tab:simulation:parameter} shows the simulation study results for the parameters, the percentage relative bias and the coverage probability of the parameters. Across the different scenarios, models exhibit varying degrees of bias in estimating parameters. Clearly, for scenario 1, Model 1 has the least bias;  for scenario 2, Model 2 has the least bias; and for scenario 3, Model 3 has the least bias. These findings underscore the critical role of selecting an appropriate model tailored to the specific data source, thereby ensuring more accurate results. The coverage probability quantifies how effectively the prediction intervals encompass the actual or true values of the outcome variable $Y$. In Scenario 1, the coverage probabilities for Model 1 were quite close to the nominal 95\%, while Model 2 and Model 3 are conservative and permissive, respectively. In Scenario 2,  the coverage probabilities for Model 2 were quite close to the nominal 95\%, while Model 1 and Model 3 are conservative. In Scenario 3,  the coverage probabilities for Model 3 were quite close to the nominal 95\%, while Model 1 and Model 2 are conservative.

Table \ref{tab:simulation:prediction} shows the simulation study results for the prediction, the bias, RMSE and the coverage probability of the prediction. In Scenario 1, we found that Model 1 has the least bias, RMSE and achieves a good coverage probability (95.32\%). This result is consistent across all the scenarios, that is, Model 2 is the best in scenario 2 and Model 3 is the best in scenario 3. These results emphasise the importance of selecting an appropriate predictive model. The predictive performance of the model depends on whether the data generation process aligns with the assumptions of the model.

\begin{table}[htp]
\centering
\caption{Summary of the result of the simulation study showing the percentage relative bias and coverage probability of the parameters}
\label{tab:simulation:parameter}
\begin{tabular}{lcccccc}
\toprule
Parameter & \multicolumn{3}{c}{Percentage relative bias} & \multicolumn{3}{c}{95\% Coverage probability} \\
\midrule
& Model 1 & Model 2 & Model 3 & Model 1 & Model 2 & Model 3 \\
\midrule
\multicolumn{7}{c}{Scenario 1} \\
\midrule
$\beta_0$  & -17.18 & 17.40 & 88.28 & 95.25\% & 95.63\% & 94.90\% \\
$\beta_1$ & -0.42 & 3.11 & 1.13 & 95.97\% & 95.99\% & 94.95\% \\
$\beta_2$ & 0.80 & -1.03 & -1.10 & 95.97\% & 95.99\% & 94.95\% \\
$\sigma^2$ & 16.40 & 34.97 & -26.91 & 95.20\% & 98.53\% & 94.90\% \\
$\phi_1$ & 13.60 & -64.19 & -99.50 & 95.83\% & 98.72\% & 94.32\% \\
$\phi_2$ & 304.08 & 1453.31 & 3.78e+09 & 95.43\% & 98.73\% & 93.98\% \\
$\phi_3$ & 598.46 & 3527.57 & 7.99e+03 & 95.39\% & 98.85\% & 93.59\% \\
\midrule
\multicolumn{7}{c}{Scenario 2} \\
\midrule
$\beta_0$ & -62.50 & -44.38 & 72.09 & 95.41\%  & 95.06\% & 95.45\% \\
$\beta_1$ & 0.05 & 0.01 & -0.43 & 95.68\% & 95.30\% & 95.78\%  \\
$\beta_2$ & -0.18 & -0.04 & -1.67 & 95.63\%  & 95.56\% & 95.48\% \\
$\sigma^2$ & 410.13 & 16.12 & 454.15  & 96.52\%  & 95.96\%  & 96.65\%\\
$\phi_1$ & 50.79 & 18.59 & -95.74 & 96.68\%  & 95.79\% & 97.36\% \\
$\phi_2$  & 498.02 & 198.40 & 4237.16 & 96.04\%  & 94.94\% & 97.56\% \\
$\phi_3$ & 557.83 & 90.31 & 1624.61 & 96.32\% & 94.94\%  & 97.12\% \\
\midrule
\multicolumn{7}{c}{Scenario 3} \\
\midrule
$\beta_0$ & -302.23 & -144.04 & -26.22 & 95.75\% & 95.66\% & 95.61\% \\
$\beta_1$ & -0.04 & 1.35 & 0.02 & 95.05\% & 95.35\% & 95.00\% \\
$\beta_2$ & 2.28 & 0.95 & 0.29 & 95.45\% & 95.42\% & 95.39\% \\
$\sigma^2$ & 1.76e+09 & 3.16e+05 & 187.29 & 96.58\% & 96.32\% & 95.36\% \\
$\phi_1$ & 3546.46 & 291.73 & 6.20 & 96.53\% & 96.42\% & 95.36\% \\
$\phi_2$  & 3442.31 & 1114.59 & 261.41 & 96.86\% & 96.54\% & 95.33\% \\
$\phi_3$ & 1.78e+04 & 5.34e+05 & 741.21 & 96.05\% & 96.52\% & 95.33\% \\
\bottomrule
\end{tabular}
\end{table}

\begin{table}[htp]
\centering
\caption{Summary of the result of the simulation study showing the bias, RMSE, and Coverage Probability for the prediction of the outcome $Y$}
\label{tab:simulation:prediction}
\begin{tabular}{lccc}
\toprule
 & {Bias} & {RMSE} & {Coverage Probability} \\
\midrule
\multicolumn{4}{c}{{Scenario 1}} \\
\midrule
{Model 1} & -3.05e-10 & 5.48e-09 & 95.32\% \\
{Model 2} & 1.86e-09 & 1.37e-08 & 96.76\% \\
{Model 3} & 9.82e-08 & 2.64e-07 & 97.78\% \\
\midrule
\multicolumn{4}{c}{{Scenario 2}} \\
\midrule
{Model 1} & -3.87e-11 & 2.91e-09 & 95.74\% \\
{Model 2} & -1.16e-11 & 2.38e-09 & 95.37\% \\
{Model 3} & -1.60e-10 & 1.10e-08 & 96.75\% \\
\midrule
\multicolumn{4}{c}{{Scenario 3}} \\
{Model 1} & 2.56e-05 & 1.65e-04 & 97.93\% \\
{Model 2} & 9.82e-08 & 2.64e-07 & 96.78\% \\
{Model 3} & -1.00e-08 & 5.83e-08 & 95.22\% \\
\bottomrule
\end{tabular}
\end{table}

\pagebreak
\section{Application of the proposed method to malaria in Mozambique} \label{sec:application}
 To illustrate the proposed approach, we fitted the nonstationary geostatistical model to map Malaria prevalence in Mozambique.  The malaria prevalence data obtained from the Malaria Atlas Project (\url{www.malariaatlas.org}) was used for this analysis. We used empirical logit prevalence as the outcome. Demographic and environmental covariates known to influence malaria transmission, such as altitude, temperature, precipitation, humidity, and proximity to water sources, are incorporated into the model. Figure \ref{fig:maps} (top-left panel) shows the predicted empirical prevalence of malaria across 447 distinct locations. This data has previously been analysed by \cite{moraga2021bayesian} and \cite{ejigu2023new}. 

 \subsection*{Demographic and environmental covariates}
 The data on temperature, precipitation and altitude were obtained from the WorldClim database (\url{www.worldclim.org}), data on the distance to inland water were derived from the Worldpop database (\url{www.worldpop.org}), and humidity data were extracted from the University of East Anglia Climatic Research Unit database (UEACRU, \url{www.cru.uea.ac.uk}) using the \textbf{R} package \textit{raster}.

\subsection*{Model formulation}
At first, a non-spatial model encompassing all available covariates was fitted. This initial model highlighted the lack of significance of precipitation while identifying the significance of other covariates, aligning with findings in \citep{ejigu2023new}. Consequently, we retained altitude, temperature, humidity, and distance from inland water as influential factors.

Subsequently, all four variables were included in the model's mean structure, with two of them integrated into the covariance function. The selection of the two variables for inclusion in the covariance function was guided by identifying the combination that yielded the lowest AIC and BIC values. We explored the three model formulations shown in Equations \eqref{eq:model1}, \eqref{eq:model2}, and \eqref{eq:model3}. Additionally, we considered the flexibility of the smoothness parameter $\kappa$ by allowing it to vary at 0.5, 1.5, and 2.5. The final model selection was made based on the one exhibiting the smallest AIC and BIC values.

Hence, our final selected model is given as follows:
\begin{equation}\label{FinalModel}
Y_i = \beta_0 + \beta_1 \textrm{Altitude} +  \beta_1 \textrm{Temperature} + \beta_2 \textrm{Humidity} + \beta_3 \textrm{Distance} + \beta_4 \textrm{Elevation} +  S(x_i, \textrm{Altitude}_i) + S(x_i, \textrm{Temperature}_i)  + Z_i,
\end{equation}
with Altitude and Temperature covariates incorporated into the covariance function, and $\kappa_1 = 1.5$, $\kappa_2 = 2.5$, and $\kappa_3 = 1.5$.

For comparison purposes, we also fitted a stationary geostatistical model (Eqn \ref{StMBG}) to the data, where only the separation distance between locations used in the spatial random effect. The stationary geostatistical model was of the form:
\begin{equation}\label{StMBG}
Y_i = \beta_0 + \beta_1 \textrm{Altitude} +  \beta_1 \textrm{Temperature} + \beta_2 \textrm{Humidity} + \beta_3 \textrm{Distance} + \beta_4 \textrm{Elevation} +  S(x_i)  + Z_i.
\end{equation}

\subsection*{Result and discussion}

Table \ref{tab:malaria_result} provides the parameter estimates and their corresponding 95\% confidence intervals for the stationary and the non-stationary models.  Altitude, temperature and humidity show significant positive associations with malaria prevalence, suggesting that higher altitudes,  elevated temperatures and high humidity contribute to increased malaria transmission based on the two models. Moreover, our result suggests that distance to inland water sources is not statistically significant in the model, challenging our initial hypothesis of elevated malaria prevalence at close distances to water bodies. This result was also observed by \cite{moraga2021bayesian}. The estimates of the scale parameter over the space are similar for the two models. The estimated value of $\tau^{2}$ under the stationary approach is slightly different from the one obtained under the non-stationary approach. This may be because the observed variability in $\tau^{2}$ is induced by altitude and temperature (in the stationary approach) which was taken into account by the specified covariance function under the non-stationary approach.

\begin{figure}[htp!]
    \centering
    \includegraphics[scale = 0.5]{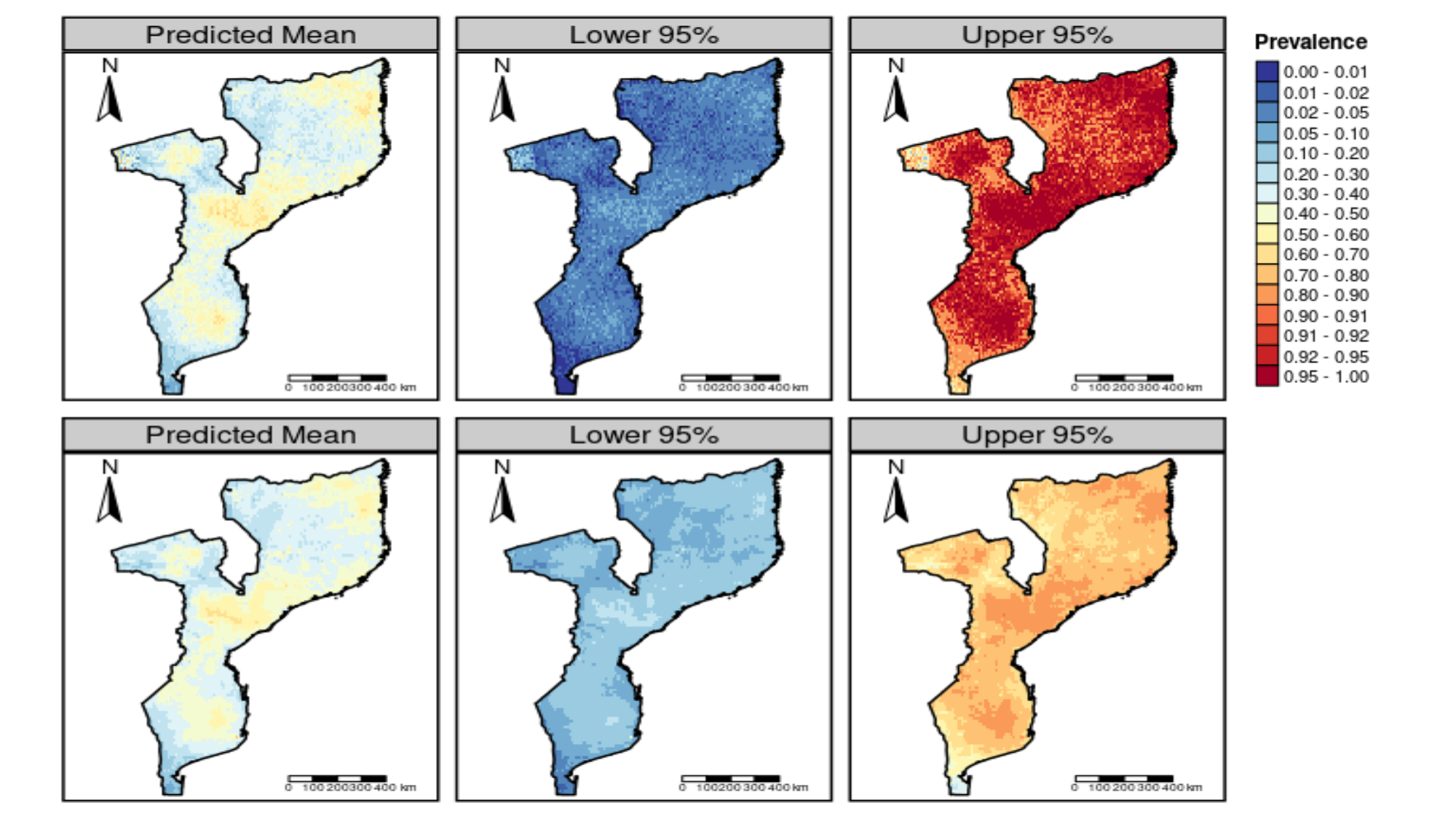}
    \caption{Map showing the results of the proposed non-stationary model is presented in the top panels, contrasting with the results from the stationary model depicted in the bottom panel. The map includes the predicted mean prevalence (left panel), lower 95\% prediction interval (middle panel) and upper 95\% prediction interval (right panel). }
    \label{fig:maps}
\end{figure}

Figure \ref{fig:maps} shows the predicted mean and upper and lower 95\% prediction intervals for the stationary (bottom panels) and the non-stationary (upper panels) models. The predicted mean prevalence of malaria is higher in the northern and central parts of the country and lower in the southern part of the country. The prevalence surfaces for the stationary and non-stationary models look similar. However, the 95\% prediction intervals for the non-stationary model appear wider than those of the stationary model, suggesting that the non-stationary model can potentially capture a broader range of uncertainties.

\begin{table}[ht]
\centering
\caption{Parameter Estimates with the corresponding 95\% confidence intervals for the stationary and the non-stationary models.}
\begin{tabular}{cccccc}
\hline
& \multicolumn{2}{c}{Non-stationary} & \multicolumn{2}{c}{Stationary} \\
\hline
Parameter & Estimate & 95\% CI & Estimate & 95\% CI \\
\hline
$\beta_0$ & -30.1355 & (-39.1782, -21.0929) & -26.0012 & (-35.2202, -16.7821) \\
$\beta_1$ & 0.0026 & (0.0016,   0.0037) & 0.0023 &  (0.0012,   0.0034) \\
$\beta_2$ & 0.5288 & (0.3628,   0.6948) & 0.4360 &  (0.2675,   0.6045) \\
$\beta_3$ & 0.1705 & (0.1112,   0.2299) & 0.1546  & (0.0939,   0.2152) \\
$\beta_4$ & 0.0139 & (-0.0001,   0.0281) & 0.0120 & (-0.0022,   0.0262) \\
$\sigma^2$ & 0.3541 & (0.1697, 0.5385) & 0.5200  & (0.1905,   0.8496) \\
$\phi_s$ & 3.1850 & (1.8273, 6.1873) & 3.7224  & (0.4359,   7.0089) \\
$\phi_\textrm{Altitude}$ & 2.4716 & (0.9596, 4.9029) & - & - \\
$\phi_\textrm{Temperature}$ & 2.1977 & (0.1146, 3.8638) & - & - \\
$\tau^2$ & 0.5368 & (0.2058, 0.8678) & 1.3579  & (0.0093,   2.7252) \\
\hline
AIC & \multicolumn{2}{c}{907.8887} & \multicolumn{2}{c}{915.3301}\\
BIC & \multicolumn{2}{c}{934.9266} & \multicolumn{2}{c}{944.9604}\\
\hline
\end{tabular}
\label{tab:malaria_result}
\end{table}

\pagebreak
\section{Discussion}
\label{sec:discussion}
This paper introduces an approach for modelling non-stationary processes within geostatistical settings by incorporating multiple covariates into the spatial process. We present three distinct methods for integrating multiple covariates into the spatial process. Results from a comprehensive simulation study underscore the importance of selecting an appropriate geostatistical model when introducing a non-stationary process. The methodology is illustrated through an analysis of malaria prevalence in Mozambique by fitting both the proposed non-stationary model and a stationary model. We conclude that the non-stationary model has made a material difference to our predictive inferences for malaria prevalence as indicated in the uncertainty. The model was implemented in R statistical software \citep{R} and the code used for the analysis can be found in \url{https://github.com/olatunjijohnson/Nonstationary_paper}.

Throughout the paper, we used two covariates to model the non-stationary process. However, this framework can be readily extended to accommodate more than two covariates (i.e., $p$) in a manner where equations \ref{eq:cov1}, \ref{eq:cov2}, and \ref{eq:cov3} can be reformulated as:

\begin{equation*}
\label{eq:cov4}
    Cov\{S(x, e_1, \ldots e_p), S(x', e'_1, \ldots e'_p)\}  = \sigma^2 \rho(x, x'; \phi) \prod_{j= 1}^{p} \rho_j(e_j, e'_j; \phi_j) 
\end{equation*}

\begin{equation*}
\label{eq:cov5}
    Cov\{S(x, e_1, \ldots e_p), S(x', e'_1, \ldots e'_p)\}  = \sigma^2 \rho_1(x, x'; \phi_1) \left(\sum_{j= 1}^{p} \rho_j(e_j, e'_j; \phi_j)\right),
\end{equation*}

and 
\begin{equation*}
\label{eq:cov6}
    Cov\{S(x, e, t), S(x', e', t')\}  = \sigma^2 \left(\rho_1(x, x'; \phi_1) + \sum_{j= 1}^{p} \rho_j(e_j, e'_j; \phi_j)\right),
\end{equation*}
respectively, where $e_1, \ldots e_p$ denotes $p$ covariates. This is particularly valuable in disease mapping when numerous environmental variables drive the disease dynamics and the underlying process is non-stationary, varying across different spatial and temporal scales. Using multiple covariates allows for a more nuanced representation of the disease-environment interactions, capturing complex relationships that may not be evident with fewer covariates.

Careful selection of covariates remains crucial to accurately representing the underlying spatial process. Additionally, these covariates can also be incorporated into the mean structure of the geostatistical model, further enriching its predictive capability. This flexibility underscores the robustness of the framework and its adaptability to diverse disease mapping scenarios, particularly in the presence of non-stationary and multi-scale drivers.

There are other ways that the covariates can be incorporated into the covariance function. \cite{risser2016nonstationary} provides an excellent review of the class of nonstationary model. One can also allow the parameters of the covariance function to be spatially varying resulting in a non-stationary process. \cite{ingebrigtsen2014spatial} shows that using stochastic differential equations for spatial modelling allows covariate information to be easily introduced in the dependence structure.

This analysis has several limitations. One limitation is that while various covariance functions can be considered, we constrain ourselves by specifying the covariance function through the fixed smoothness parameter $\kappa$ of the Matérn covariance function. Future work could involve allowing this parameter to be estimated from the data, although it's essential to consider that, as noted by \cite{zhang2004inconsistent}, estimating $\kappa$ from the data typically requires a substantial amount of data.

Future work includes extending the current framework to account for anisotropy and account for directionality. To achieve this, Mahalanobis distance between locations, as suggested by \cite{schmidt2011considering}, may be considered. Also, there is an opportunity to extend the model to non-linear settings where the relationship between the outcome and the covariates is modelled through smooth functions.

% \section*{Ethics approval}
% Not applicable. 
\section*{Competing interests}
The authors declare that there is no conflict of interest regarding the publication of this article.
% \section*{Acknowledgements}
% OJ and EG express gratitude for the support provided by the London Mathematical Society, which sponsored EG for a research visit to the UK (grant reference 52214).
\section*{Funding}
This research did not receive any specific grant from funding agencies in the public, commercial, or not-for-profit sectors. 
\section*{Data accessibility}

% \section*{Authors’ contributions}

\bibliographystyle{elsarticle-harv}

%\bibliography{cas-refs}

\begin{thebibliography}{18}
\expandafter\ifx\csname natexlab\endcsname\relax\def\natexlab#1{#1}\fi
\providecommand{\url}[1]{\texttt{#1}}
\providecommand{\href}[2]{#2}
\providecommand{\path}[1]{#1}
\providecommand{\DOIprefix}{doi:}
\providecommand{\ArXivprefix}{arXiv:}
\providecommand{\URLprefix}{URL: }
\providecommand{\Pubmedprefix}{pmid:}
\providecommand{\doi}[1]{\href{http://dx.doi.org/#1}{\path{#1}}}
\providecommand{\Pubmed}[1]{\href{pmid:#1}{\path{#1}}}
\providecommand{\bibinfo}[2]{#2}
\ifx\xfnm\relax \def\xfnm[#1]{\unskip,\space#1}\fi
%Type = Article
\bibitem[{Alexandra~M. et~al.(2011)Alexandra~M., Peter and
  Anthony}]{Schmidt2011}
\bibinfo{author}{Alexandra~M., S.}, \bibinfo{author}{Peter, G.},
  \bibinfo{author}{Anthony, O.}, \bibinfo{year}{2011}.
\newblock \bibinfo{title}{Considering covariates in the covariance structure of
  spatial processes}.
\newblock \bibinfo{journal}{Environmetrics} \bibinfo{volume}{22},
  \bibinfo{pages}{487--500}.
%Type = Article
\bibitem[{Amoah et~al.(2022)Amoah, Fronterre, Johnson, Dejene, Seife, Negussu,
  Bakhtiari, Harding-Esch, Giorgi, Solomon et~al.}]{amoah2022model}
\bibinfo{author}{Amoah, B.}, \bibinfo{author}{Fronterre, C.},
  \bibinfo{author}{Johnson, O.}, \bibinfo{author}{Dejene, M.},
  \bibinfo{author}{Seife, F.}, \bibinfo{author}{Negussu, N.},
  \bibinfo{author}{Bakhtiari, A.}, \bibinfo{author}{Harding-Esch, E.M.},
  \bibinfo{author}{Giorgi, E.}, \bibinfo{author}{Solomon, A.W.}, et~al.,
  \bibinfo{year}{2022}.
\newblock \bibinfo{title}{Model-based geostatistics enables more precise
  estimates of neglected tropical-disease prevalence in elimination settings:
  mapping trachoma prevalence in ethiopia}.
\newblock \bibinfo{journal}{International Journal of Epidemiology}
  \bibinfo{volume}{51}, \bibinfo{pages}{468--478}.
%Type = Article
\bibitem[{Burton et~al.(2006)Burton, Altman, Royston and
  Holder}]{burton2006design}
\bibinfo{author}{Burton, A.}, \bibinfo{author}{Altman, D.G.},
  \bibinfo{author}{Royston, P.}, \bibinfo{author}{Holder, R.L.},
  \bibinfo{year}{2006}.
\newblock \bibinfo{title}{The design of simulation studies in medical
  statistics}.
\newblock \bibinfo{journal}{Statistics in medicine} \bibinfo{volume}{25},
  \bibinfo{pages}{4279--4292}.
%Type = Article
\bibitem[{Christopher~J. and Mark~J.(2006)}]{Paciorek2006}
\bibinfo{author}{Christopher~J., P.}, \bibinfo{author}{Mark~J., S.},
  \bibinfo{year}{2006}.
\newblock \bibinfo{title}{Spatial modelling using a new class of nonstationary
  covariance functions}.
\newblock \bibinfo{journal}{Environmetrics} \bibinfo{volume}{17},
  \bibinfo{pages}{485--506}.
%Type = Article
\bibitem[{Diggle et~al.(1998)Diggle, Tawn and Moyeed}]{diggle1998model}
\bibinfo{author}{Diggle, P.J.}, \bibinfo{author}{Tawn, J.A.},
  \bibinfo{author}{Moyeed, R.A.}, \bibinfo{year}{1998}.
\newblock \bibinfo{title}{Model-based geostatistics}.
\newblock \bibinfo{journal}{Journal of the Royal Statistical Society Series C:
  Applied Statistics} \bibinfo{volume}{47}, \bibinfo{pages}{299--350}.
%Type = Article
\bibitem[{Ejigu and Moraga(2023)}]{ejigu2023new}
\bibinfo{author}{Ejigu, B.A.}, \bibinfo{author}{Moraga, P.},
  \bibinfo{year}{2023}.
\newblock \bibinfo{title}{A new way of analyzing malaria data: A non-stationary
  geostatistical modeling approach} .
%Type = Article
\bibitem[{Ejigu et~al.(2020)Ejigu, Wencheko, Moraga and
  Giorgi}]{ejigu2020geostatistical}
\bibinfo{author}{Ejigu, B.A.}, \bibinfo{author}{Wencheko, E.},
  \bibinfo{author}{Moraga, P.}, \bibinfo{author}{Giorgi, E.},
  \bibinfo{year}{2020}.
\newblock \bibinfo{title}{Geostatistical methods for modelling non-stationary
  patterns in disease risk}.
\newblock \bibinfo{journal}{Spatial Statistics} \bibinfo{volume}{35},
  \bibinfo{pages}{100397}.
%Type = Article
\bibitem[{Higdon(1998)}]{higdon1998process}
\bibinfo{author}{Higdon, D.}, \bibinfo{year}{1998}.
\newblock \bibinfo{title}{A process-convolution approach to modelling
  temperatures in the north atlantic ocean}.
\newblock \bibinfo{journal}{Environmental and Ecological Statistics}
  \bibinfo{volume}{5}, \bibinfo{pages}{173--190}.
%Type = Article
\bibitem[{Ingebrigtsen et~al.(2014)Ingebrigtsen, Lindgren and
  Steinsland}]{ingebrigtsen2014spatial}
\bibinfo{author}{Ingebrigtsen, R.}, \bibinfo{author}{Lindgren, F.},
  \bibinfo{author}{Steinsland, I.}, \bibinfo{year}{2014}.
\newblock \bibinfo{title}{Spatial models with explanatory variables in the
  dependence structure}.
\newblock \bibinfo{journal}{Spatial Statistics} \bibinfo{volume}{8},
  \bibinfo{pages}{20--38}.
%Type = Article
\bibitem[{Johnson et~al.(2022)Johnson, Giorgi, Fronterr{\`e}, Amoah, Atsame,
  Ella, Biamonte, Ogoussan, Hundley, Gass et~al.}]{johnson2022geostatistical}
\bibinfo{author}{Johnson, O.}, \bibinfo{author}{Giorgi, E.},
  \bibinfo{author}{Fronterr{\`e}, C.}, \bibinfo{author}{Amoah, B.},
  \bibinfo{author}{Atsame, J.}, \bibinfo{author}{Ella, S.N.},
  \bibinfo{author}{Biamonte, M.}, \bibinfo{author}{Ogoussan, K.},
  \bibinfo{author}{Hundley, L.}, \bibinfo{author}{Gass, K.}, et~al.,
  \bibinfo{year}{2022}.
\newblock \bibinfo{title}{Geostatistical modelling enables efficient safety
  assessment for mass drug administration with ivermectin in loa loa endemic
  areas through a combined antibody and loascope testing strategy for
  elimination of onchocerciasis}.
\newblock \bibinfo{journal}{PLoS Neglected Tropical Diseases}
  \bibinfo{volume}{16}, \bibinfo{pages}{e0010189}.
%Type = Article
\bibitem[{Mat{\'e}rn(1960)}]{matern1960spatial}
\bibinfo{author}{Mat{\'e}rn, B.}, \bibinfo{year}{1960}.
\newblock \bibinfo{title}{Spatial variation, technical report}.
\newblock \bibinfo{journal}{Statens Skogsforsningsinstitut, Stockholm} .
%Type = Article
\bibitem[{Mogaji et~al.(2022)Mogaji, Johnson, Adigun, Adekunle, Bankole,
  Dedeke, Bada and Ekpo}]{mogaji2022estimating}
\bibinfo{author}{Mogaji, H.O.}, \bibinfo{author}{Johnson, O.O.},
  \bibinfo{author}{Adigun, A.B.}, \bibinfo{author}{Adekunle, O.N.},
  \bibinfo{author}{Bankole, S.}, \bibinfo{author}{Dedeke, G.A.},
  \bibinfo{author}{Bada, B.S.}, \bibinfo{author}{Ekpo, U.F.},
  \bibinfo{year}{2022}.
\newblock \bibinfo{title}{Estimating the population at risk with soil
  transmitted helminthiasis and annual drug requirements for preventive
  chemotherapy in ogun state, nigeria}.
\newblock \bibinfo{journal}{Scientific Reports} \bibinfo{volume}{12},
  \bibinfo{pages}{2027}.
%Type = Article
\bibitem[{Moraga et~al.(2021)Moraga, Dean, Inoue, Morawiecki, Noureen and
  Wang}]{moraga2021bayesian}
\bibinfo{author}{Moraga, P.}, \bibinfo{author}{Dean, C.},
  \bibinfo{author}{Inoue, J.}, \bibinfo{author}{Morawiecki, P.},
  \bibinfo{author}{Noureen, S.R.}, \bibinfo{author}{Wang, F.},
  \bibinfo{year}{2021}.
\newblock \bibinfo{title}{Bayesian spatial modelling of geostatistical data
  using inla and spde methods: A case study predicting malaria risk in
  mozambique}.
\newblock \bibinfo{journal}{Spatial and Spatio-temporal Epidemiology}
  \bibinfo{volume}{39}, \bibinfo{pages}{100440}.
%Type = Manual
\bibitem[{{R Core Team}(2023)}]{R}
\bibinfo{author}{{R Core Team}}, \bibinfo{year}{2023}.
\newblock \bibinfo{title}{R: A Language and Environment for Statistical
  Computing}.
\newblock \bibinfo{organization}{R Foundation for Statistical Computing}.
  \bibinfo{address}{Vienna, Austria}.
\newblock \URLprefix \url{https://www.R-project.org/}.
%Type = Article
\bibitem[{Risser(2016)}]{risser2016nonstationary}
\bibinfo{author}{Risser, M.D.}, \bibinfo{year}{2016}.
\newblock \bibinfo{title}{Nonstationary spatial modeling, with emphasis on
  process convolution and covariate-driven approaches}.
\newblock \bibinfo{journal}{arXiv preprint arXiv:1610.02447} .
%Type = Article
\bibitem[{Sampson and Guttorp(1992)}]{Sampson1992}
\bibinfo{author}{Sampson, P.D.}, \bibinfo{author}{Guttorp, P.},
  \bibinfo{year}{1992}.
\newblock \bibinfo{title}{Nonparametric estimation of nonstationary spatial
  covariance structure}.
\newblock \bibinfo{journal}{Journal of the American Statistical Association}
  \bibinfo{volume}{87}, \bibinfo{pages}{108--119}.
%Type = Article
\bibitem[{Schmidt et~al.(2011)Schmidt, Guttorp and
  O'Hagan}]{schmidt2011considering}
\bibinfo{author}{Schmidt, A.M.}, \bibinfo{author}{Guttorp, P.},
  \bibinfo{author}{O'Hagan, A.}, \bibinfo{year}{2011}.
\newblock \bibinfo{title}{Considering covariates in the covariance structure of
  spatial processes}.
\newblock \bibinfo{journal}{Environmetrics} \bibinfo{volume}{22},
  \bibinfo{pages}{487--500}.
%Type = Article
\bibitem[{Zhang(2004)}]{zhang2004inconsistent}
\bibinfo{author}{Zhang, H.}, \bibinfo{year}{2004}.
\newblock \bibinfo{title}{Inconsistent estimation and asymptotically equal
  interpolations in model-based geostatistics}.
\newblock \bibinfo{journal}{Journal of the American Statistical Association}
  \bibinfo{volume}{99}, \bibinfo{pages}{250--261}.

\end{thebibliography}

\end{document}